\def\simlt{\lower.5ex\hbox{$\; \buildrel < \over \sim \;$}}
\def\simgt{\lower.5ex\hbox{$\; \buildrel > \over \sim \;$}}
\RequirePackage{lineno} 
 \documentclass[preprint2]{emulateapj}
 \usepackage{graphicx}
 \usepackage{mathtools} 

\newcommand{\myemail}{mrline@ucsc.edu}
\slugcomment{Submitted to the Astrophysical Journal }
\shorttitle{Patchy Cloud Transmission}
\shortauthors{Line \& Parmentier}

\begin{document}
\title{The Influence of Non-Uniform Cloud Cover on Transit Transmission Spectra}
\author{Michael R. Line}
\affil{NASA Hubble Postdoctoral Fellow}
\affil{NASA Ames Research Center}
\affil{Bay Area Environmental Research Institute}
\affil{School of Earth and Space Exploration, Arizona State University}
\author{Vivien Parmentier}
\affil{NASA Sagan Postdoctoral Fellow}
\affil{Department of Astronomy \& Astrophysics, University of California-Santa Cruz}
\affil{Lunar \& Planetary Laboratory, University of Arizona}

\email{mrline@ucsc.edu}
\altaffiltext{1}{Correspondence to be directed to \myemail}

\begin{abstract}
We model the impact of non-uniform cloud cover on transit transmission spectra.  Patchy clouds exist in nearly every solar system atmosphere, brown dwarfs, and transiting exoplanets. Our major findings suggest that fractional cloud coverage can exactly mimic high {  mean molecular weight} atmospheres and vice-versa over certain wavelength regions, in particular, over the Hubble Space Telescope (HST) Wide Field Camera 3 (WFC3) bandpass (1.1-1.7 $\mu$m).  We also find that patchy cloud coverage exhibits a signature that is different from uniform global clouds.  {  Furthermore, we explain analytically why the ``patchy cloud-high mean molecular weight" degeneracy exists.} We also explore the degeneracy of non-uniform cloud coverage in atmospheric retrievals on both synthetic and real planets.  We find from retrievals on a synthetic solar composition hot Jupiter with patchy clouds and a cloud free high mean molecular weight warm Neptune, that both cloud free high mean molecular weight atmospheres and partially cloudy atmospheres can explain the data equally well.  Another key find is that the HST WFC3 transit transmission spectra of two well observed objects, the hot Jupiter HD189733b and the warm Neptune HAT-P-11b, can be explained well by solar composition atmospheres with patchy clouds without the need to invoke high mean molecular weight or global clouds. The degeneracy between high molecular weight and solar composition partially cloudy atmospheres can be broken by observing the molecular Rayleigh scattering differences between the two. Furthermore,  the signature of partially cloudy limbs also appears as a $\sim$100 ppm residual in the ingress and egress of the transit light curves, provided the transit timing is known to seconds.

\end{abstract}
\section{Introduction}

Clouds are ubiquitous in our solar system, brown dwarfs, and extra solar planets. They often complicate the interpretation of spectra due to our ignorance of their properties. This can strongly inhibit our ability to place precision constraints on other atmospheric properties like temperatures and abundances (Ackerman \& Marley 2001; Kirpatrick 2005; Cushing et al. 2008; Burrows et al. 2006; Howe \& Burrows 2012; Morley et al. 2013;15; Kriedberg et al. 2014a;15; Vahidinia et al. 2014; Knutson et al. 2014; Benneke 2015; Marley \& Robinson 2014). In particular, they strongly impact transit transmission spectra, because the longer optical path lengths permit a greater sensitivity to trace cloud species (Fortney et al. 2005).  To date, only one-dimensional, limb averaged cloud models have been used to interpret the transmission spectra of exoplanets (Fortney et al. 2010; Howe \& Burrows 2012; Morley et al. 2013;15; Kriedberg et al. 2014a;15; Fraine et al. 2014; Benneke 2015).  In this investigation we show that the presence of {\it inhomogeneous} clouds along the terminators of transiting exoplanets can strongly influence our interpretation of current transit transmission spectra.

Inhomogeneous clouds are common in our solar system over a broad range of bulk planetary properties (terrestrial, jovian, cold, hot etc.).  Meridional atmospheric cells combined with vertical temperature gradients create the banded clouds seen on Earth and on Jupiter;  warm, humid parcels of gas rise from one latitude in the atmosphere where they then cross the dew point temperature, at which point condensates form. The now dry air then sinks back down leading to clear skies at a different latitude.  Similar processes are presumed to happen in Brown dwarf atmospheres (Burgasser et al. 2002; Marley et al. 2010; Zhang \& Showman 2014), where observations are strongly suggestive of non-uniform cloud cover (Buenzli et al. 2012;14; Radigan et al. 2012;14; Crossfield et al. 2014;  Apai et al. 2013; Metchev et al. 2015).

In tidally locked exoplanets, two mechanisms have been identified to form inhomogeneous clouds. First, the meridional circulation can transport condensate material from the equatorial regions to the polar regions, a phenomenon suggested to be common in hot Jupiter and sub-Neptune planets (Parmentier et al. 2013, Charnay et al. 2015a). The second mechanism is specific to the hottest of the tidally locked planets. The intense and inhomogeneous stellar irradiation they receive creates a large day-night temperature contrast which drives a strong west-to-east atmospheric circulation pattern. This atmospheric circulation, dominated by a super-rotating equatorial jet, advects thermal energy eastward leading to a strong west-to-east terminator temperature gradient of several hundreds of degrees. This eastward shift in temperature was first predicted by Showman and Guillot (2002) and since  observed in a a number of hot Jupiter's (Knutson et al. 2007;2009;2012; Crossfield et al. 2010; Cowan et al. 2012; Lewis et al. 2013; Stevenson et al. 2014). Such a large horizontal temperature gradient can lead to longitudinally varying cloud cover as numerous condensable species can be in a condensed state on the west limb but gaseous in the east limb.  Recent phase curve observations in visible light from the Kepler spacecraft are strongly suggestive of inhomogeneous dayside cloud coverage, with cloudy western daysides and clear eastern daysides (Demory et al. 2013, Shporer \& Hu 2015; Hu et al. 2015; Esteves et al. 2015; Parmentier et al. 2015, submitted), very much in agreement with expectations.

In the remainder of the paper we describe how a non-uniform cloud cover along the planetary terminator can influence the observed transit transmission spectra and how failing to account for non-uniform cloud cover can bias molecular abundance determinations. In \S\ref{sec:Methods} we illustrate the basic idea and describe the impact that non-uniform terminator cloud cover can have on transit transmission spectra. {  \S\ref{sec:analytic} reviews, analytically, the basic mechanisms that control the shape of transit transmission spectra and the role of non-uniform terminator cloud cover.}   In \S\ref{sec:Results} we show quantitatively, via atmospheric retrievals, how non-uniform cloud cover can impact water abundance determinations on both synthetic data and two well observed planets, the hot-Jupiter HD189733b and warm Neptune HAT-P-11b.  In \S \ref{sec:Light_Curve} we show how non-uniform cloud cover can present itself as residuals in transit light curves. Finally we summarize our findings, caveats, and discuss implications.

\section{Basic Concept and Impact's on Transit Transmission Spectra}\label{sec:Methods}
%%%%%%%%%%%%%figure1%%%%%%%%%%%%%%%%%%%%
\begin{figure}[h]
%\begin{center}
\includegraphics[width=0.48\textwidth, angle=0]{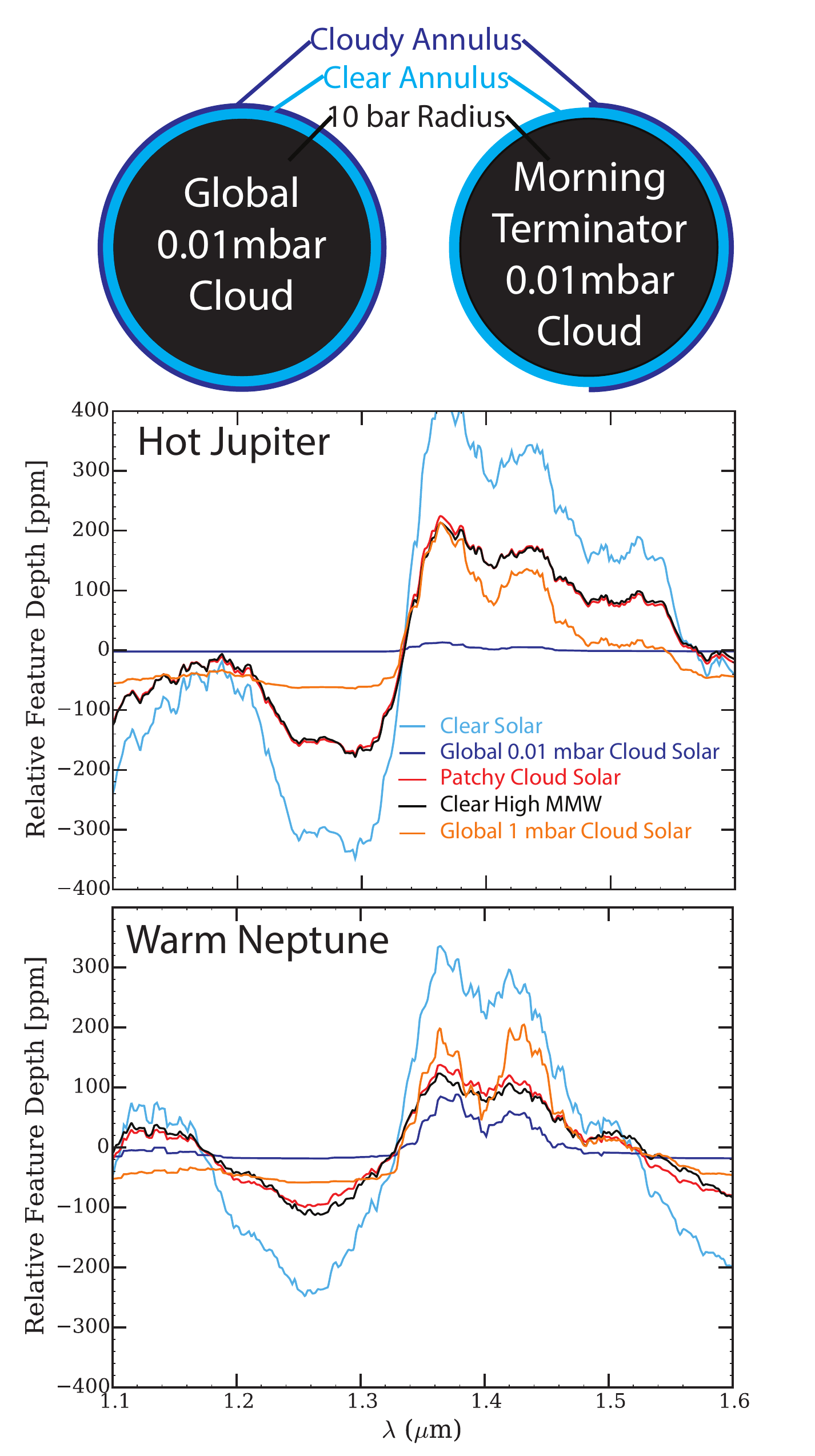}
%\end{center}
     \caption{ \label{fig:Figure1} Impact of non-uniform cloud cover on a hot-Jupiter and warm-Neptune transit transmission spectrum.  The cartoon at the top illustrates the relative change in radius (to scale for the hot Jupiter) due to the the clear and cloudy annuli. In the bottom two panels we compare representative spectra of different commonly encountered atmospheric scenarios for a hot Jupiter (middle) and warm Neptune (bottom) (see Table \ref{tab:table1}) with partial cloudy spectra .     The spectra are offset to have zero mean.  Note the near identical match between the patchy cloud and high {  mean molecular weight (mmw)} spectra.}  
\end{figure} 
%%%%%%%%%%%%figure1%%%%%%%%%%%%%%%%%%%%
We use a transit transmission forward model (Line et al. 2013; Swain, Line, \& Deroo 2014; Kreidberg et al. 2014b;15) to generate a variety of spectra over the HST WFC3 bandpass to illustrate the basic concept.  The model numerically solves the equations described in Brown (2001) and Tinetti et al. (2012).  The inputs are the scale height isothermal temperature (T), {  the planetary radius at 10 bars}, the opaque ({  in the limb geometry}) gray cloud top pressure (P$_{c}$), a terminator cloud fraction ($f$), and the gas abundances.  For simplicity we include only water as the gaseous absorber as it has been the only molecule robustly detected over the WFC3 bandpass (e.g., Deming et al. 2013; Kreidberg et al. 2014; 2015).  The remaining ``filler'' gas is assumed to be a mixture of molecular hydrogen and helium in solar proportions.  Clouds are modeled rather simplistically assuming that no light is transmitted through the atmosphere at pressures deeper than P$_c$.  Non-uniform cloudy transmission spectra are computed via a linear combination of a globally clear atmosphere and globally cloudy atmosphere (similar to Marley et al. 2010; Morley et al. 2014 for brown dwarf emission spectra) using the following:
\begin{equation}\label{eq:equation1}
\alpha_{\lambda, f}=f\alpha_{\lambda, cloudy}+(1-f)\alpha_{\lambda, clear}
\end{equation}
where $\alpha_{\lambda, f}$ is the wavelength dependent eclipse depth for cloud fraction $f$, $\alpha_{\lambda, cloudy}$ and $\alpha_{\lambda, clear}$ are the wavelength dependent eclipse depths for a globally cloudy and globally clear atmosphere, respectively.  This is effectively a ``2-dimensional" model as we don't consider variations along the tangent rays (unlike in Fortney 2010; Burrows et al. 2010). Figure \ref{fig:Figure1} illustrates the basic concept for two representative planets, a hot Jupiter and a warm Neptune. The necessary planet and atmosphere parameters are shown in Table \ref{tab:table1}.

%%%%%%%%%%%%%%%%%%%%%%%%%%Table1-Parameters%%%%%%%%%%%%%%%%%%%%
\begin{table*}
\centering
\caption{\label{tab:table1} Parameters used to generate the simulated representative planet atmospheres shown in Figure \ref{fig:Figure1}}
%\resizebox{\textwidth}{!}{%
\begin{tabular}{lcc}
%\footnote{\tiny{aa}}
\hline
\hline
\cline{1-2}
Parameter & Hot Jupiter & Warm Neptune    \\
\hline
Planet Radius at 10 bars ($R_{Jupiter}$)   & 1.25  &  0.38 \\
Stellar Radius ($R_{sun}$)  & 1.15 &  0.46\\
Planet gravity ($m/s$) &10  & 12.6  \\
Isothermal Temperature ($K$) & 1500 & 700  \\
Solar Water Mixing Ratio\footnote{Thermochemical equilibrium abundance of water at the isothermal temperature and 1 $mbar$ using solar elemental abundances} & 4.5$\times 10^{-4}$  &  6.3$\times 10^{-4}$\\
High Metallicity Water Mixing Ratio\footnote{These are the water abundances required to adequately match the partially cloudy spectra--the ``$\times$Solar" is the ratio between this water abundance and the solar water abundance, and not the elemental abundances}  & 1.5$\times 10^{-1}$  (250$\times$Solar)  & 3.0$\times 10^{-1}$ (500$\times$ Solar) \\
High Metallicity Mean Molecular Weight (a.m.u)\footnote{the same mean molecular weights can be achieved with $\sim$125$\times$ and $\sim$250$\times$ solar metallicity, respectively, when taking into account the contributions to the mean molecular weight from other molecules}&4.6 & 7 \\
Global Cloud Top Pressure ($mbar$)\footnote{For the clear atmospheres, this is set to the deepest pressure layer in the model grid, 30 $bars$}&  1   & 1 \\
Patch Cloud Top Pressure ($mbar$) & 0.01  & 0.01 \\
Patchy Cloud Fraction\footnote{Fraction of terminator covered in the 0.01 $mbar$ clouds. This is set to 1 for the global clouds, and 0 for clear atmospheres} &  0.5	&  0.6 \\
\hline
\end{tabular}
\end{table*}
%%%%%%%%%%%%%%%%%%%%%%%%%%Table1%%%%%%%%%%%%%%%%%%%%%%%%%%%

For this setup we use a relatively high altitude opaque cloud in order to substantially flatten the spectra, a reasonable cloud top pressure in lieu of recently observed flat transmission spectra (Kreidberg et al. 2014, Knutson et al. 2014, Morley et al. 2012;2015) and GCM tracer studies (Parmentier et al. 2013, Charnay et al. 2015b).  We see that the non-uniform cloud cover damps the spectral features as it is an average of a near flat line and a clear atmosphere. Certainly different cloud distributions (e.g., a cloudy northern hemisphere, clear southern hemisphere) could have the same terminator cloud fraction. For a given cloud fraction, we would not be able to determine the spatial distribution of clouds along the terminator, however this degeneracy may be broken through high precision transit light curves (\S \ref{sec:Light_Curve}).  

The {\it shape} of the non-uniform cloud cover spectra is much different than a globally uniform deeper cloud (red curve vs. green curve).  A globally uniform deeper cloud has a flatter shape  $\le$1.32 $\mu$m and a somewhat deeper trough near the peak at 1.4 $\mu$m, and a steeper slope $>$1.36 $\mu$m.  The most interesting result from this exercise is the near perfect match between the non-uniform cloud cover (red) and the {  high mean molecular weight\footnote{  by high mean molecular weight we mean high enough to begin to shrink spectral features, which in the scenarios in this paper begins around a molecular weight of 2.8 (see also Benneke \& Seager 2013). Such a molecular weight for typical solar elemental proportions in hydrogen dominated atmospheres occurs near metallicities of $\sim 60 \times$ solar.} (typically assumed to be due to high metalicity)} clear spectrum (black) for both planet setups.  The degeneracy between {\it uniform} clouds and {  high mean molecular weight} has already been investigated in a variety of planets (Benneke \& Seager 2013; Knutson et al. 2014; Fraine et al. 2014).  The non-uniform cloud degeneracy is different as the shape and the amplitude of the higher {  mean molecular weight} spectral features in the WFC3 bandpass can be almost exactly reproduced by the non-uniform cloud cover, but less so by a globally uniform cloud.  In the next section we describe, analytically, what controls the shape of spectral features in transmission and why non-uniform cloud cover exactly mimics high mean molecular weight atmospheres.
%In order to more rigorously explore the degeneracies introduced by non-uniform cloud cover, we turn to atmospheric retrievals on realistic observational scenarios.

{ 
\section{What Controls the Shape of Transmission Spectra}\label{sec:analytic}
Here we discuss the variables that control the shape of spectral features in transit transmission spectra within the simple analytic formulation presented by Lecavelier Des Etangs et al. (2008). 
\subsection{One Absorber}
The original Lecavelier Des Etangs et al. (2008) formulation (see their equation 1) computes the equivalent wavelength dependent sharp occulting disk radius, $R_{p}+z_{\lambda}$ (where $R_{p}$ is wavelength independent) for a single absorber with
\begin{equation}\label{eq:zed}
z_{\lambda}=\frac{k_{b}T}{\mu g}\ln\left(\xi \sigma_{\lambda} \frac{1}{\sqrt{k_{b}T\mu g}}\beta \right)
\end{equation}
and
\begin{equation}
\beta=\frac{P_{0}}{\tau_{eq}}\sqrt{2\pi R_{p}}
\end{equation}
where $k_{b}$ is Boltzman's constant, $T$ is temperature, $\mu$ is the atmospheric mean molecular weight, $g$ is gravity, $\xi$ is the volume mixing ratio of a particular absorber with cross section $\sigma_{\lambda}$, $P_{0}$ is the reference pressure at ``0 altitude", and $\tau_{eq}$ is the optical depth required to produce and equivalent opaque disk. The wavelength dependent eclipse depth $(\alpha_{\lambda}$) can then be given with $((R_{p}+z_{\lambda})/R_{star})^2$. If we assume that the atmospheric thickness is much smaller than the planet radius, we can approximate the eclipse depth as
\begin{equation}\label{eq:alpha}
\alpha_{\lambda}\approx\left(\frac{R_{p}}{R_{star}}\right)^2+\frac{2R_{p}z_{\lambda}}{R_{star}^2}
\end{equation}
We define the spectral shape as the wavelength dependent slope, $d\alpha_{\lambda}/d\lambda$. Differentiating (\ref{eq:alpha}) with respect to $\lambda$ gives
\begin{equation}\label{eq:shape_one}
\frac{d\alpha_{\lambda}}{d\lambda}=\frac{2R_{p}}{R_{star}^2}H\frac{d\ln(\sigma_{\lambda})}{d\lambda}
\end{equation}
where $\frac{k_{b}T}{\mu g}$ is the scale height,$H$, similar to equation (2) in Lecavelier Des Etangs et al. (2008). From this we see that for a given planet ($R_{p}$ and $R_{star}$) the spectral shape depends on the wavelength dependence of the absorption cross section and the scale height.  

Note how the volume mixing ratio of the absorber, $\xi$, does not appear in this relation.   This suggests that given a single spectrally dominant trace species over some wavelength range, the spectral shape is independent of the molecular abundance. As a consequence it is not possible to measure the molecular abundance (Benneke \& Seager 2012; Griffith 2013), unless using the small pressure dependence of absorption cross­ section (de Wit \& Seager 2013).  This breaks down when the abundance of the trace species is large enough to impact the mean molecular weight or when an additional absorber is present.
\subsection{Multiple Absorbers}
We now ask how the spectral shape depends on contributions from multiple absorbers. This will be important for understanding the contribution of clouds to the spectral shape.  We begin by generalizing the above equations for $N$ absorbers, $i$, with
\begin{equation}
z_{\lambda}=H\ln\left(\frac{1}{\sqrt{k_{b}T\mu g}}\beta \sum_{i}^N\xi_{i} \sigma_{\lambda,i} \right)
\end{equation}
and the spectral shape,
\begin{equation}\label{eq:shape_eq}
\frac{d\alpha_{\lambda}}{d\lambda}=\frac{2R_{p}}{R_{star}^2}H\frac{1}{\sum_{i}^N\xi_{i} \sigma_{\lambda,i}}\sum_{i}^N\xi_{i} \frac{d\sigma_{\lambda,i}}{d\lambda}
\end{equation}

For the remainder of the analysis lets consider only two opacity sources with absorption cross section $\sigma_{1}$ and abundance $\xi_{1}$ for absorber 1 and $\sigma_{2}$ and abundance $\xi_{2}$ for absorber 2.  Equation (\ref{eq:shape_eq}), after some manipulation, then becomes
\begin{equation}\label{eq:shape_two}
\frac{d\alpha_{\lambda}}{d\lambda}=\frac{2R_{p}}{R_{star}^2}H\frac{1}{1+\frac{\xi_{2}\sigma_{\lambda,2}}{\xi_{1}\sigma_{\lambda,1}}}\left(\frac{d\ln\sigma_{\lambda,1}}{d\lambda}+\frac{\xi_{2}\sigma_{\lambda,2}}{\xi_{1}\sigma_{\lambda,1}}\frac{d\ln\sigma_{\lambda,2}}{d\lambda}\right).
\end{equation}
This shows that, in addition to the scale height modulating the spectral slope, we now have to consider the shape modulation due to the relative contributions of the different absorbers. Provided there are no other dependencies, only the ratios of abundances can be measured (Benneke \& Seager 2012).  In a clear, solar composition hot jupiter atmosphere the dominant absorber contributions will come from water, molecular Rayleigh scattering, and the H$_2$-H$_2$/He collision induced absorption (CIA). If a feature is attributed to CIA and another to water, then the water­-to-H$_2$/He abundance ratio can be measured. That ratio combined with the idea that all mixing ratios have to sum to unity permits the absolute abundance determinations (Benneke \& Seager 2012). This is effectively what sets a lower bound on the water abundance.  Identifying spectral regions where the cross-section contrast is minimized (e.g., $\sigma_{\lambda,1}\approx\sigma_{\lambda,2}$) would permit a stronger sensitivity to the trace absorber as the slopes in equation (\ref{eq:shape_two}) are more balanced.  If a feature is attributed to clouds and another to water, then there is a strong uncertainty on the water abundance because of our lack of priors on the cloud properties. CIA and grey clouds are distinguishable in the WFC3 bandpass because the CIA has a non-gray opacity structure (see Figure 6 in Line et al. 2013).

Now we assume the two absorbers are water and a gray cloud uniformly distributed through the atmosphere\footnote{{  This analytic formulation assumes vertical homogeneity in the absorbers and thus does not permit the ``hard cloud top pressure" parameterization used in many spectral interpretations, including all of the numerical simulations in this investigation. However, the results in this analytic discussion are appropriate as one can find an ``equivalent" hard cloud top pressure that mimics the gray uniform absorber. Increasing the uniform cloud opacity has the same effect as decreasing the hard cloud top pressure.}}. The gray cloud cross-section, by definition, has zero spectral slope so equation (\ref{eq:shape_two}) reduces to
\begin{equation}\label{eq:shape_three}
\frac{d\alpha_{\lambda}}{d\lambda}=\frac{1}{1+\frac{\xi_{cld}\sigma_{\lambda,cld}}{\xi_{H2O}\sigma_{\lambda,H2O}}}\frac{2R_{p}}{R_{star}^2}H\frac{d\ln(\sigma_{\lambda,H2O})}{d\lambda}.
\end{equation}
When the cloud opacity is low ($\xi_{H2O}\sigma_{H2O}>>\xi_{cld}\sigma_{cld}$), then equation (\ref{eq:shape_three}) reduces to equation (\ref{eq:shape_one}). If the opposite is true then the spectral modulation approaches zero.  Note how the pre-factor multiplying the wavelength dependent cross section is also wavelength dependent, unlike equation (\ref{eq:shape_one}). This means that an atmosphere with a gray global cloud with an opacity comparable to that of water (e.g., such that the spectral features are muted but not completely flat) can not be exactly mimicked by a clear, smaller scale height atmosphere.

\subsection{Patchy Clouds}
Now we consider how the spectral shape is modulated by non-uniform terminator cloud cover assuming the cloudy portion of the terminator is dominated by the cloud opacity.   Combining equation (\ref{eq:equation1}) with equation (\ref{eq:shape_one}) (the clear atmosphere spectral shape) and the assumption that the cloud opacity dominates over the molecular absorber opacity on the cloudy terminator  (such that the spectral slope is zero) we obtain
\begin{equation}\label{eq:shape_patchy}
\frac{d\alpha_{\lambda}}{d\lambda}=(1-f)\frac{2R_{p}}{R_{star}^2}H\frac{d\ln(\sigma_{\lambda,H2O})}{d\lambda}.
\end{equation}
This is effectively the same as equation (\ref{eq:shape_one}) for a clear atmosphere except that the pre-factor that scales the wavelength dependent cross-section now depends on the terminator cloud fraction, $f$.  From this, it can be readily seen that there is a degeneracy between scale height and terminator cloud fraction, and why a high mean-molecular-weight atmosphere can identically mimic a low mean molecular weight atmosphere with partial terminator cloud coverage.  Note how with the partial cloudy atmosphere there is no wavelength dependence to the pre-factor, in contrast to the global cloud (equation (\ref{eq:shape_three})).  This suggests, in theory, that partially cloudy and fully cloudy atmospheres are therefore distinguishable.

One may ask why we are sensitive to the mean molecular weight as opposed to the temperature in this cloud-scale height degeneracy.  Technically, the scale height depends in exactly the same way on both the molecular weight and temperature (we assume gravity is well determined). Doubling the mean molecular weight has the same effect as halving the temperature on the scale height, hence spectral slope. However, the temperature not only affects the scale height but the strength of the absorption cross sections. It is the temperature dependence of the absorption cross-sections that permits us to disentangle temperature from molecular weight.   Furthermore our {\it a priori} knowledge of the planetary temperature from energy balance arguments is greater than that of the mean molecular weight (Benneke \& Seager 2012). That is, it's more physically plausible to increase the mean molecular weight of the atmosphere by an order of magnitude, than the limb temperature.  In the next section we quantitatively explore the patchy cloud degeneracy within an atmospheric retrieval framework.}
\section{Impact on Atmospheric Retrievals}\label{sec:Results}
\subsection{Simulated Data}\label{sec:Synthetic}
\begin{figure*}[h]
\begin{center}
\includegraphics[width=1\textwidth, angle=0]{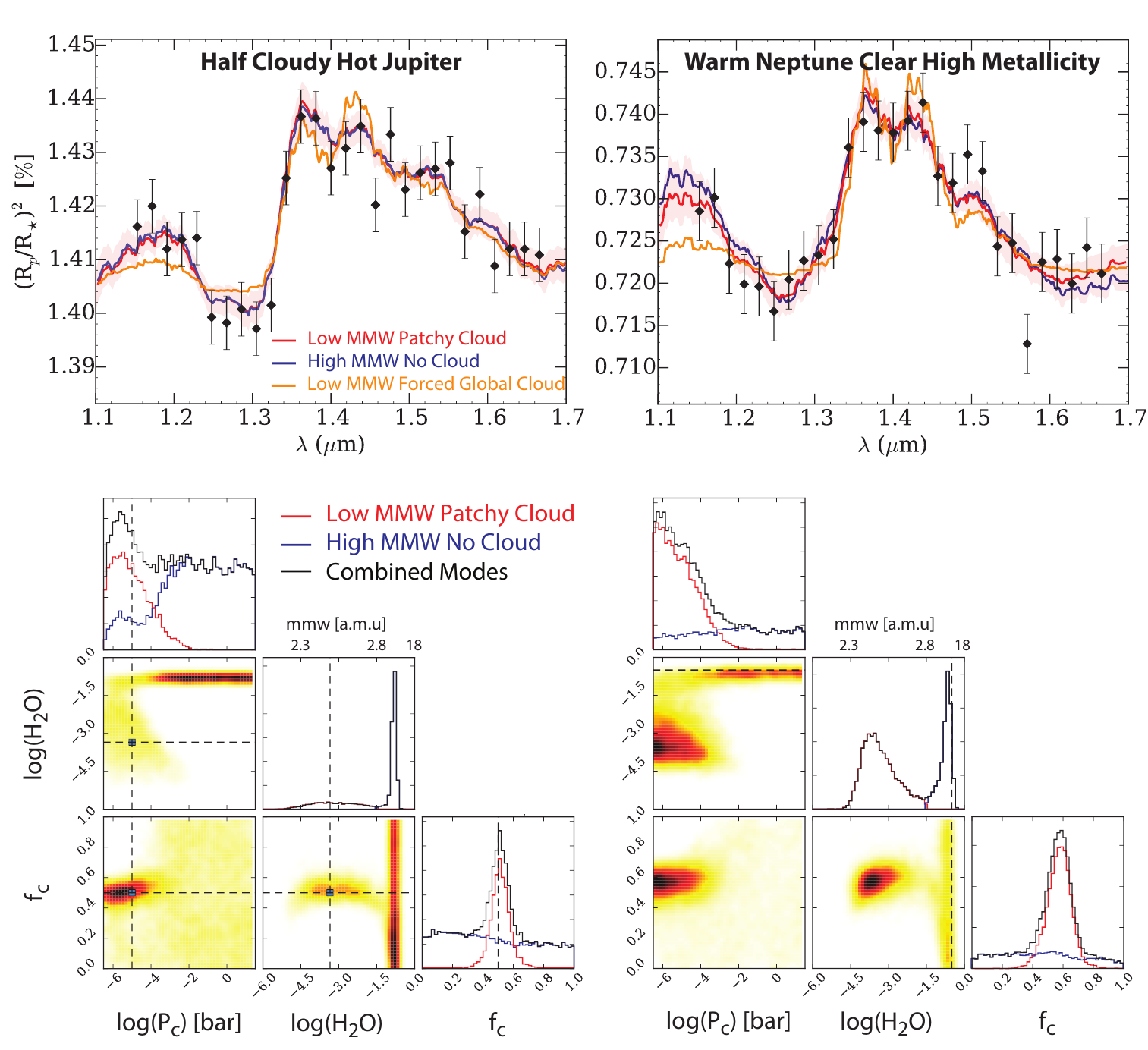}
\end{center}
     \caption{ \label{fig:Figure2}  Synthetic retrieval results. The top row shows the synthetic data (diamonds with error bars) for the hot-Jupiter half-cloud half-clear scenario (left) and the warm-Neptune high mean molecular weight scenario (right). For each the 2 sigma spread in spectra in derived from both modes is shown in light red. The median spectra drawn from the high and low water abundance modes only are shown in blue and red, respectively. The orange spectra are the median  for a retrieval in which there is a globally uniform cloud and and a prior upper limit restricting the high water abundance mode. The spectral modulation from one  scale height is 150 ppm in the solar composition hot Jupiter, and 32 ppm for the high molecular weight warm Neptune.   The bottom row shows the posterior probability distributions for the interesting parameters (water abundance, cloud top pressure, and terminator cloud fraction).  The black histograms are the resultant 1D marginalized distributions for each parameter.  The red and blue curves result from drawing samples only from the low or high water ({  high molecular weight}) modes, respectively (where the cut is at logH$_2$O=-1.5, mean molecular weight of 2.8).   The dashed lines show the true values for each scenario (the true cloud and cloud fraction for the warm Neptune case are on the edges at 1.5 and 0 respectively).    Note that both scenarios can be fit well by either a high {  mean molecular weight} or a partial cloud cover at approximately {  solar composition molecular weight}.  In the high {  mean molecular weight} mode the cloud location and coverage are largely ill informed (blue histograms). }
\end{figure*}

We generate two representative synthetic WFC3 observations and explore the degeneracies introduced by partial cloud cover.  The first is a 1500 K solar composition (450 ppm of water) hot-Jupiter with a 50\% terminator cloud fraction at 0.01 mbar (same as the scenario described in \S\ref{sec:Methods} and Table \ref{tab:table1}). We randomly draw a simulated set of observations with 50 ppm error bars and 0.02 $\mu$m resolution (e.g., Kreidberg et al. 2014b). The second scenario is representative of the warm-Neptune class of planets (e.g., GJ436b, HAT-P-11b). It is a 700 K high metallicity (500$\times$ the solar water abundance in thermochemical equilibrium--mean molecular weight of $\sim$7 amu) cloud free atmosphere (cloud fraction 0, cloud depth at 30 bars) with 35 ppm error bars (e.g., Kreidberg et a. 2014a).  Figure \ref{fig:Figure2} (top row) shows the simulated spectra for each scenario (black diamonds with error bars).  Again,  water is the only trace gas absorber.  Similar feature sizes in the high mean molecular weight scenarios can be obtained with lower metallicities when taking into account the contribution to the mean molecular weight due to other heavier molecules.

We employ {\it PyMultiNest}\footnote{http://johannesbuchner.github.io/PyMultiNest/index.html} (Buchner et al. 2014), a python wrapper to the commonly used {\it MultiNest} (Feroz et al. 2009) nested sampling algorithm, to explore the parameter degeneracies.  Nested sampling algorithms are well suited for exploring multi-modal and high correlated parameter spaces (e.g., see Benneke \& Seager 2013 for application to super earth transmission spectra) as well as the efficient computation of the bayesian evidence.  Figure \ref{fig:Figure2} summarizes the retrieved results. The bottom row shows the retrieved probability distributions for the interesting parameters (the scale heigh temperature and 10 bar radius are not shown for clarity). The water abundance histograms for both scenarios clearly have two modes.  The first mode is the near solar water abundance mode (solar mean molecular weight of 2.3) requiring non-uniform cloud cover. The second is a high water abundance {  (corresponding to a mean molecular weight of $>$2.8)} mode requiring no cloud at all (e.g., only a lower pressure, high altitude limit).  

To further explore this bi-modality we isolate these two modes by drawing samples from the full posterior with water abundances that fall in either the high (blue) or low (red) water models (the cut being made at a log(H$_2$O) value of -1.5).  The median spectra from each of these modes are shown in the top row of Figure \ref{fig:Figure2}.   In both observational scenarios we find that the low water mode requires fractional cloud coverage with approximately half of the terminator covered in a high altitude (P $<$ 0.1 mbar) clouds.  In the high water abundance {  (high mean molecular weight)} mode there is no longer a strong sensitivity to the location of the cloud (and correspondingly the cloud fraction), as the large {  limb} optical depth obscures the presence of any cloud down to $\sim$0.1 mbar pressures . 

In order to be more quantitative, we compute the log of the Bayes factor ($lnB$) for the modes, a metric for quantitatively comparing two competing models (e.g., Cornish \& Littenberg 2007).  For the hot-Jupiter scenario we obtain an $lnB$ of 0.7 (considered negligible on the Jeffery's scale, Trotta 2008) suggesting we cannot confidently distinguish the two modes, though the preference is slightly weighted towards the high water/{  mean molecular weight} abundance mode. For the high mean molecular weight warm-Neptune scenario the preference for one mode over the other is also negligible ($lnB$=0.5).   These conclusions are unsurprising given the median spectra (top row, Figure \ref{fig:Figure2}) for each mode are nearly identical (as expected given Figure \ref{fig:Figure1}).

For completeness we also ask the question of how well these spectra can be explained with a simple uniform cloud cover model with solar-like composition of water.  To do this we perform a separate retrieval where we place an upper bound on the water abundance prior at logH$_2$O=-1.5. This effectively eliminates the possibility of the retrieval searching out the high {  mean molecular weight} mode.  Figure \ref{fig:Figure2} (top row) shows the median spectra (in orange) from this set up.  These spectra are substantially different than the patchy cloud or high {  mean molecular weight} median spectra, with flatter features $<$1.3 $\mu$m.   The bayesian evidence suggests a strong preference of the patchy cloud {\it or} high metallicity scenarios over the globally cloudy solar-like composition  ($lnB$ of $\sim$7 for the hot Jupiter and $\sim$5 for the warm-Neptune). This suggests that our synthetic observational set ups can distinguish non-uniform cloud cover (or high {  molecular weight}) from a solar composition ({  low mean molecular weight}) atmosphere with uniform cloud cover .

%what sets H2O lower bound?? Well it's the fact that H2-He *are* spectrally active via the CIA--when H2O low enough features from H2He begin to show-to make matters
%worse, as H2O abundance low, "effective" pressure at which the trans spec is probing is deeper so P^2 H2 opacity even stronger relative to water cross section
%also pressure broadening of H2O lines as well...

%what sets H2O upper bound?? definitely mean molecular weight.

%-where H2O cross section is weaker, between the bands.

%HOT JUPITER
%Full all modes: -21.61 
%low Mode Evidence:-22.23
%high mode Evidence:-21.54
%cloudy low mode:-28.526

%WARM NEPTUNE
%Full all modes: -20.77
%low Mode Evidence: -20.645
%high mode Evidence:-20.133
%cloudy low mode:-25.414

%%%%%%%%%%%%%figure2%%%%%%%%%%%%%%%%%%%%
%%%%%%%%%%%%figure2%%%%%%%%%%%%%%%%%%%%

\subsection{Application to Real Planets: HD189733b \& HAT-P-11b}
We investigate the possibility of non-uniform cloud cover on two observed planets: the well studied, canonical hot-Jupiter, HD189733b, and the recently characterized warm-Neptune, HAT-P-11b. 

We briefly summarize the previous conclusions inferred from the HST WFC3 transit observations of these planets.  HD189733b was observed with WFC3 in spatial scan mode by McCullugh et al. (2014). Madhusudhan et al. (2014) presented the first extensive retrieval analysis of the HD189733b HST WFC3 transmission spectra. They found, under the {\em assumption} of a cloud free atmosphere, that a low water abundance was required to explain the spectra due to the relatively muted features.  They concluded that a high carbon-to-oxygen ratio was required to explain the apparent depletion of water.

A follow up retrieval study of 8 transmission spectra by Benneke (2015) also looked at HD189733b. He compared the different abundance results (within a 1D self-consistent framework) that one obtains under the assumption of various degrees of cloud parameterizations, from the simple ``Rayleigh Haze+opaque cloud deck" to parameterized profiles, assumed cloud compositions, and particle size distributions.  He found, unsurprisingly, that with the addition of more free cloud parameters, that the uncertainties in the abundances increased to a wider range of C/O ratios, but still included C/O $>$ 1 within the 99.7$\%$ confidence interval.

HAT-P-11b (Bakos et al. 2010), a $\sim$900 K warm-Neptune, was the first object in the warm-Neptune category to have claimed a water detection in transmission (Fraine et al. 2014).  They concluded that a predominantly clear high metallicity ($\sim$few tens-300$\times$ solar at 1-sigma confidence) atmosphere was most likely needed to explain the observed modulation. 

\begin{figure*}[h]
\begin{center}
\includegraphics[width=0.85\textwidth, angle=0]{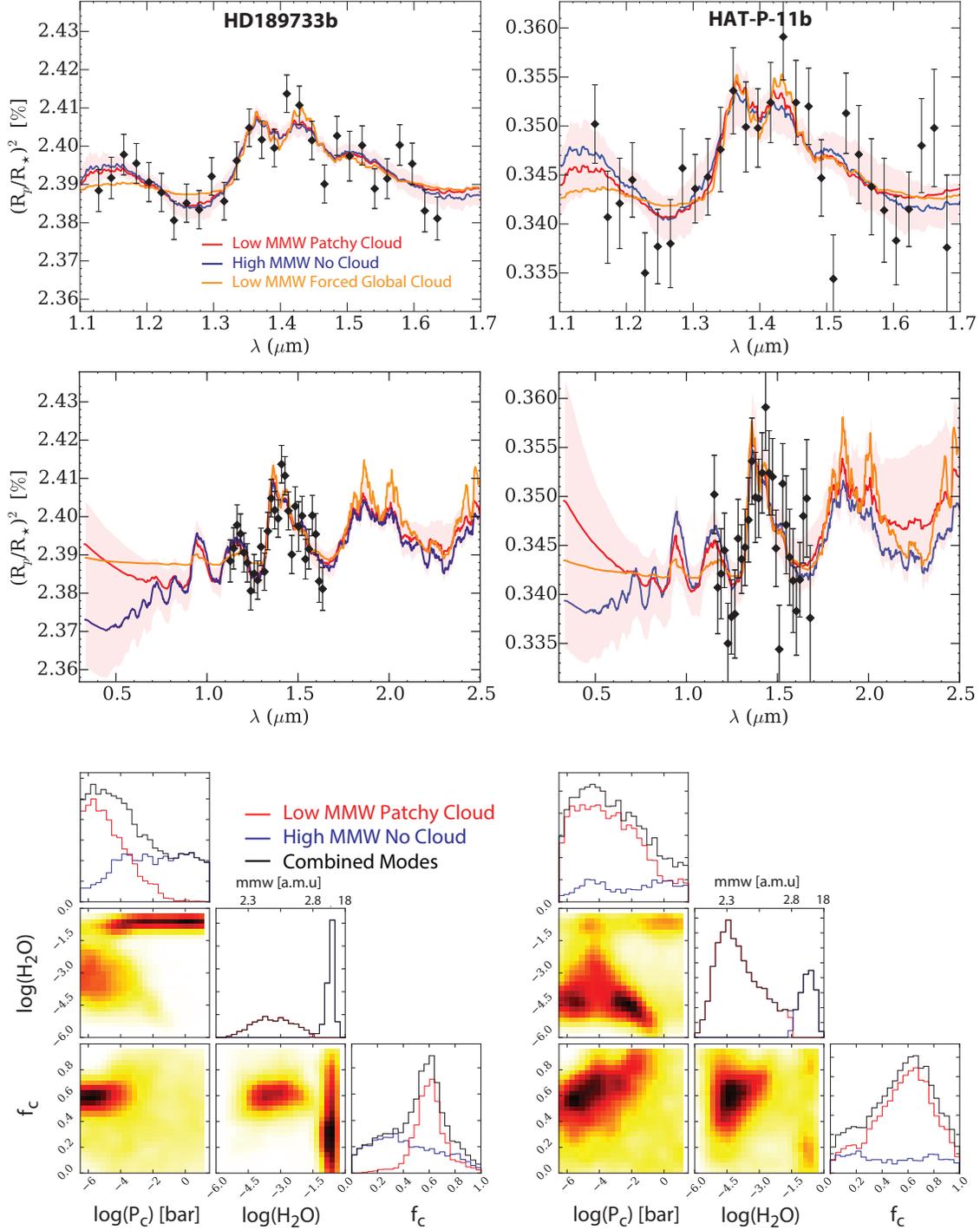}
\end{center}
     \caption{ \label{fig:Figure3}  Retrieval Results for HD189733b and HAT-P-11b. {  We have included only water, H$_2$ Rayleigh scattering, H$_2$-H$_2$/He CIA, and gray clouds as absorbers.}  The top row shows the WFC3 data (diamonds with error bars) for HD189733b (left) and HAT-P-11b (right). For each, the 2 sigma spread in spectra drawn from both mode is shown in light red. The median spectra drawn from the high and low water abundance modes are shown in blue and red, respectively. The orange curve in each is the median spectrum for a retrieval in which there is a globally uniform cloud and and a prior upper limit restricting the high water abundance mode.  The middle row shows a zoomed out version from 0.3 - 2.5 $\mu$m. While all three scenarios are consistent with the HST WFC3 data, they diverge significantly at shorter wavelengths.  The bottom row shows the posterior probability distributions for the interesting parameters  (water abundance, cloud top pressure, and terminator cloud fraction).  The black histograms are the resultant 1D marginalized distributions for each parameter.  The red and blue histograms result from drawing samples only from the low {  (mean molecular weights of 2.30 - 2.37 amu at 68\% confidence for HD189733b and 2.30- 2.30 amu for HAT-P-11b )} or high {  (mean molecular weights of 4.52 - 6.92 amu at 68\% confidence for HD189733b,  and 3.38 - 7.94 amu for HAT-P-11b)} water modes, respectively (where the cut is at logH$_2$O=-1.5, {  mean molecular weight=2.8 amu}).     The dashed lines show the true values for each scenario.  Note that both scenarios can be fit well by either {  high mean molecular weight} or by fractional cloud cover at approximately solar abundances.  In the {  high mean molecular weight} mode the cloud location and coverage are largely ill informed (blue histograms) {  due to the lack of sensitivity to those parameters at high water optical depths}.}  
\end{figure*}

We add a different interpretation to these two objects within the patchy cloud framework by performing identically the same analysis as was done on the synthetic data in \S\ref{sec:Synthetic}.  Figure \ref{fig:Figure3} summarizes these results. As with the synthetic cases, for both planets, we again find two possible solutions: the {  high mean molecular weigh mode (resulting from the high water abundance) and the solar composition low mean molecular weight } mode. 

The correlations in Figure \ref{fig:Figure3} are rich in information. For both planets, the red histograms, corresponding to the ``patchy cloud" mode, always encompasses a cloud fraction of unity (global cloud coverage), albeit at low probability suggesting that some ``clearness" is favored.  Also in this patchy cloud mode, the cloud top pressures are required to be relatively low (high altitudes) for HD189733b (P$_{c}<\sim1mbar$) with a cloud fraction between 0.5 - 0.7 (1-sigma width). The results are similar for HAT-P-11b, but less constrained due to the lower feature signal-to-noise.  A more noticeable correlation between cloud top pressure and cloud fraction appears in HAT-P-11b. As the cloud top pressure decreases (higher altitudes) the cloud fraction must also decrease, otherwise the water feature damps too much.   

Looking at the log(H$_{2}$O) vs. log(P$_{c}$) panels for both planets we find that as the cloud moves to deeper pressures the water abundance must decrease in order to preserve the low amplitude features.  {  These solutions correspond to the cloud free ``high C/O" solution found by Madhusudhan et al. (2014) for HD189733b; less favorable fits in our models. The water features are damped when the water abundance is low because of the relative weighting of the water opacity to the hydrogen CIA (equation (\ref{eq:shape_two}) with $\xi_{1}\sigma_{1}$ corresponding to water with $\xi_{2}\sigma_{2}$ corresponding to the CIA). This damping has a different behavior than a cloud as the CIA has a shape to it and is not spectrally flat. This increased CIA-to-water opacity due to the decreasing water abundance also explains why we can obtain a lower limit to the water mixing ratio constraint. A lower water abundance would result in pure CIA features.}

 For both objects we find that as the log(H$_2$O) increases to $\sim$-1.5 (3\%) the noticeable change in the atmospheric mean molecular weight {  (2.8 amu)} begins to damp the spectral features (e.g., same as in Figure 6 of Benneke \& Seager 2013). Once the water abundance is high enough (blue histograms--high water abundance/mean molecular weight mode), the tangent optical depths reach unity at low enough pressures to be at altitudes above where the clouds appear to have an impact. This permits the cloud top pressure and hence, cloud fraction, to take on nearly any value (except for really low cloud top pressures).

%Discuss plausibility of high mmsw for HD189 given M&R
%Discuss high C/O no cloud mode--madhu

Again, to be more quantitative, we undergo the same Bayes factor analysis as in \S\ref{sec:Synthetic}.   For HD189733b, a weak to moderate favorability of the high metallicity/{  mean molecular weight} mode over the partly cloudy solar water mode ($lnB$ = 1.9) is found. Both the partly cloud and high metallicity mode are moderately favored over a solar composition with global cloud cover ($lnB$=2.8). For HAT-P-11b all three scenarios are indistinguishable ($lnB$ $<\sim$1 ).  This is an important point: The {  signal-to-noise} in the current WFC3 data for both HD189733b and HAT-P-11b cannot definitively distinguish the difference between cloud free high metallicity atmospheres, solar composition patchy cloud atmospheres, or solar composition globally cloudy atmospheres.  {  As shown with the numerical examples, and analytically, one cannot distinguish between high mean molecular weights and low mean molecular weight with fractional cloud cover over the WFC3 bandpass alone. However, high enough feature signal-to-noise ratios are enough to break the degeneracy between the low mean molecular weight partially cloudy or high mean molecular weight clear atmospheres from globally cloudy deeper clouds. }

If we look at prediction for other wavelengths (middle row Figure \ref{fig:Figure3}) based on the fits to the WFC3 wavelengths we can see some divergence in the different scenarios, especially at wavelengths $<$1$\mu$m. {  This is because of the relative weighting between the water opacity and the gaseous Rayleigh scattering. In the high mean molecular weight atmosphere the water absorption dominates over the molecular Rayleigh scattering, where-as in the patchy cloud solar-abundance scenario, the rayleigh scattering dominates over the water opacity (see the analytic discussion in $\S$3.)} While shorter wavelengths can break the degeneracy between these three scenarios, the addition of a uniform high altitude Rayleigh scattering haze layer, {  or additional opacity sources such as alkali metals, metal hydrides, and oxides} could potentially thwart our ability to do so.  HD189733b has been observed from 0.3 -1.0 $\mu$m (Sing et al. 2008) and shows a strong slope with a large eclipse depth--much larger than the models presented here.  The amplitude of the slope could either be do to scattering hazes (Sing et al. 2008) or to a {  lesser effect}, star spots (McCullough et al. 2014). Strong scattering hazes in the optical {  could possibly} make breaking the degeneracy between the three scenarios difficult. {    Though, despite these complications it is apparent that the visible wavelengths offer the greatest chance of breaking this degeneracy.}

HAT-P-11b was also observed in the Kepler bandpass. However, the broad band nature of those observations prevent a detailed characterization of the short wavelength slopes and uncertainties in the relative offsets between datasets observed at different epochs under different instrumental and astrophysical conditions may also muddle our ability to disentangle the three scenarios. Longer wavelength spectra covering gases that present themselves more strongly at higher metallicities (e.g, CO , CO$_2$) should be able to break the high metallicity versus patchy cloud scenario. In the next section we look to transit light curves as an observational signature for non-uniform cloud cover.

%Evidences
%HAT-P-11b
%   -FULL: -23.95
%   -global cloud: 
%   -global cloud forced solar: -24.51
%   -forced high Z mode: -23.26
%   -forced solar: -24.52

%HD189
%  -FULL:  -23.92
%  -global cloud: 
%  -global cloud forced solar: -26.34
% -forced high Z mode: -22.58
% -forced solar mode: -24.51

%%%%%%%%%%%%%figure2%%%%%%%%%%%%%%%%%%%%
%%%%%%%%%%%%figure2%%%%%%%%%%%%%%%%%%%%
\section{Impact on Transit Light Curves}\label{sec:Light_Curve}

 \begin{figure}
\includegraphics[width=\linewidth]{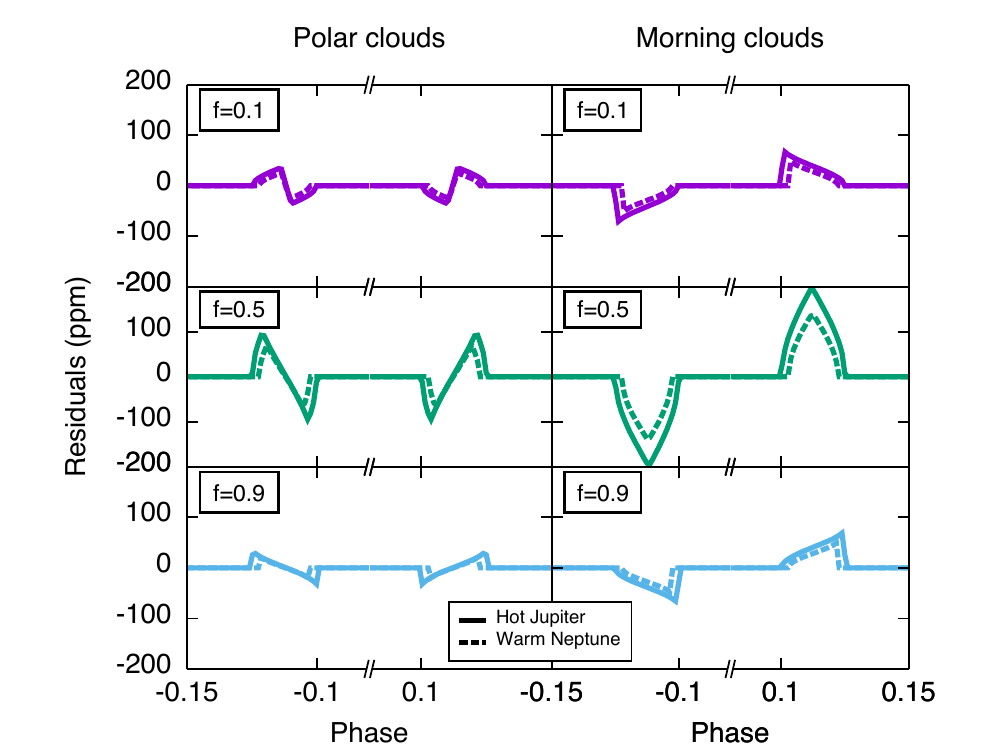}
\caption{\label{fig:Figure4}Relative difference between the transit light curve of a partially cloudy planet and the transit light curve of a spherical planet with the same apparent area. The left column is for a planet with banded clouds where the equator is cloudless and the poles are cloudy. The right column is for a planet with clouds on the morning terminator and clear on the evening terminator. The different rows are for different cloud fraction at the limb. Solid lines are for the synthetic hot Jupiter scenario and dotted lines for the synthetic warm Neptune scenario. The difference in the effective opaque radius between the cloudy and clear terminators is of five scale heights. The light curve residuals are directly proportional to this difference. We assumed no limb darkening and computed the cloudy and cloudless light curves with the \textit{occultsmall.f} subroutine of Mandel \& Agol 2002.}
\end{figure}

During the eclipse phase of a transit, all the observed quantities are averaged over the planetary limb, leading to degeneracies in the transit spectrum. The shape of the ingress and the egress of the transit, however, is determined by the shape of the planetary limb and can potentially be used to constrain the cloud distribution over the planet limb and break the degeneracies between partial cloudiness and high mean molecular weight atmospheres.

We construct the transit light curve of a partially cloudy planet by combining the transit light curve of a planet with a clear atmosphere ($\psi_{\rm clear}$) with that of a cloudy atmosphere ($\psi_{\rm cloudy}$). The transit light curve of the annulus of cloud can be constructed by subtracting the transit light curve of a clear planet from that of the cloudy planet : $\psi_{\rm ring}=\psi_{\rm cloudy}-\psi_{\rm clear}$. We consider two extreme cases, corresponding to the two cloud forming scenarios on tidally locked planets discussed in the introduction : the first is a planet with morning clouds only (i.e. no dependence on latitude but a dependence on longitude) and the second is a planet with polar clouds and a clear equator (i.e. no dependance on longitude but a dependence in latitude). In both cases we call $\alpha_{\rm c}$ the angle between the planet equator and the beginning of the cloudy evening terminator, counting from the morning side. The cloud coverage is therefore $f=1-\alpha_{c}/\pi$ for the morning cloud case and $f=1-2\alpha_{c}/\pi$ for the polar cloud case. We call $\alpha$ the angle between the morning side of the planet equator and the intersection between the stellar and the planetary limb :
\begin{equation}
\alpha=acos\left(\frac{p}{2z}+\frac{z}{2p}-\frac{1}{2pz}\right)
\end{equation}
where $p$ is the ratio of the planetary radius to the stellar radius and $z$ is the apparent distance between the center of the star and the center of the planet divided by the stellar radius, as in Mandel \& Agol (2002). $\alpha$ is defined only during the ingress and the egress of the planet and goes from $\pi$ to $0$ during both ingress and egress. 

We now assume that the planet transits the stellar equator and that the planet is small compared to the star (e.g. the stellar luminosity is considered constant over the apparent size of the planet). Outside of transit, the transit light curve is $\psi=1$. During the full phase of the transit, the light curve is $\psi=\psi_{\rm clear}+f\psi_{\rm ring}$. During the ingress the phase curve can be split into several parts. For morning clouds we have :
\begin{equation}
%\begin{spreadlines}{pt}
\psi=\left\{
\begin{aligned} 
&\psi_{\rm clear}+\psi_{\rm ring}\,, &\text{for }& \alpha>\alpha_{\rm c}\\
&\psi_{\rm clear}+\psi_{\rm ring}\frac{\pi-\alpha_{\rm c}}{\pi-\alpha}\, , &\text{for }&\alpha<\alpha_{\rm c}\, ,\\ 
%&\psi_{\rm clear}, &\text{for }& \pi-\alpha_{\rm c}<\alpha<\alpha_{\rm }<\pi\\
\end{aligned}
\right.
%\end{spreadlines}
\end{equation}
during the ingress and
\begin{equation}
%\begin{spreadlines}{pt}
\psi=\left\{
\begin{aligned} 
&\psi_{\rm clear}+\psi_{\rm ring}\frac{\alpha-\alpha_{\rm c}}{\alpha}\,, &\text{for }& \alpha>\alpha_{\rm c}\\
&\psi_{\rm clear}, &\text{for }&\alpha<\alpha_{\rm c}\, ,\\
%&\psi_{\rm clear}, &\text{for }& \pi-\alpha_{\rm c}<\alpha<\alpha_{\rm }<\pi\\
\end{aligned}
\right.
%\end{spreadlines}
\end{equation}
during the egress. For the polar cloud we have:
%For polar clouds they are :
\begin{equation}
%\begin{spreadlines}{pt}
\psi=\left\{
\begin{aligned} 
&\psi_{\rm clear}\,, &\text{for }& \alpha>\pi-\alpha_{\rm c}\\
&\psi_{\rm clear}+\psi_{\rm ring}\frac{\pi-\alpha-\alpha_{\rm c}}{\pi-\alpha}\,,&\text{for }&\alpha_{\rm c}<\alpha<\pi-\alpha_{\rm c}\\
&\psi_{\rm clear}+\psi_{\rm ring}\frac{\pi-2\alpha_{\rm c}}{\pi-\alpha}\,,&\text{for }&\alpha<\alpha_{\rm c}\, ,
%&\psi_{\rm clear}+\psi_{\rm ring}\frac{\pi-\alpha_{\rm c}}{\pi-\alpha}\, , &\text{for }&\alpha<\alpha_{\rm c}\\
%&\psi_{\rm clear}, &\text{for }& \pi-\alpha_{\rm c}<\alpha<\alpha_{\rm }<\pi\\
\end{aligned}
\right.
%\end{spreadlines}
\end{equation}
during the ingress and 
\begin{equation}
%\begin{spreadlines}{pt}
\psi=\left\{
\begin{aligned} 
&\psi_{\rm clear}+\psi_{\rm ring}\frac{\pi-2\alpha_{\rm c}}{\alpha}\,, &\text{for }& \alpha>\pi-\alpha_{\rm c}\\
&\psi_{\rm clear}+\psi_{\rm ring}\frac{\alpha-\alpha_{\rm c}}{\alpha}\,,&\text{for }&\alpha_{\rm c}<\alpha<\pi-\alpha_{\rm c}\\
&\psi_{\rm clear}\,,&\text{for }&\alpha<\alpha_{\rm c}\, ,
%&\psi_{\rm clear}+\psi_{\rm ring}\frac{\pi-\alpha_{\rm c}}{\pi-\alpha}\, , &\text{for }&\alpha<\alpha_{\rm c}\\
%&\psi_{\rm clear}, &\text{for }& \pi-\alpha_{\rm c}<\alpha<\alpha_{\rm }<\pi\\
\end{aligned}
\right.
%\end{spreadlines}
\end{equation}
during the egress. The transit light curve of a partially cloudy planet can be computed by combining in the manner above, the transit light curve of a cloudy and a cloudless planet, each computed separately via the IDL routines provided by Mandel \& Agol (2002).

Shown in Figure~\ref{fig:Figure4}, the transit light curve of a partially cloudy planet will present residuals when modeled with a single uniform light curve. The amplitude of these residuals are $\sim100\,\rm ppm$. For these illustrative calculations, we assume that the effective radius of the cloudy planet is 5 scale heights larger than that of the clear, corresponding to clouds having an opacity $\approx150$ times higher than the gas opacity in the bandpass of observation (equation 1, Lecavelier Des Etangs et al. 2008 ). The residuals scale linearly with the difference in effective radius between the clear and cloudy atmospheres,  thus with the difference between the natural log of the cloud and clear opacities. The signal should therefore be largest in bandpasses where the gaseous opacities are small.

As seen in Figure \ref{fig:Figure4}, the amplitude of the residuals is similar for the hot Jupiter and warm Neptune cases. This is because the residuals are proportional to the ratio of the areas of the cloudy annulus to the stellar disk. In our case the smaller scale height of our fiducial warm-Neptune model is compensated by the smaller radius of the star it transits (Table \ref{tab:table1}).

The shape of the residuals strongly depends on the distribution of clouds. Whereas the ingress and egress are symmetric in the case of polar clouds, they are anti-symmetric in the case of morning clouds. A more complex picture is expected for more complex cloud distributions and could become degenerate with other effects such as planet oblateness (e.g., Carter et al. 2010).

Usually when interpreting transit light curves, both the planet and orbital parameters are fit simultaneously. {  This causes an additional degeneracy between the morning terminator cloud model and the center-of-transit time for a standard uniform limb model. We explored this degeneracy through some additional light curve models.  We found that the light curve residuals in Figure \ref{fig:Figure4} are reduced by a factor of ten given a shift in the transit timing of $\sim 10^{-5}$ and $\sim 10^{-4}$ days for a planet with a 1 and 10 scale height difference in radius between the clear and cloudy half's of the planet, respectively}.  We thus recommend high precision transit timing to infer the presence of non-uniform cloud cover.

\section{Discussion \& Conclusions}\label{sec:Conclusions}
We have demonstrated, {  both numerically and analytically}, that non-uniform, or ``patchy" clouds" can impact the interpretation of transit transmission spectra.  Albedo phase curve measurements (Demory et al. 2013, Hu et al. 2015, Esteves et al. 2015) and general circulation models with tracers (e.g., Parmentier et al. 2013; Charnay et al. 2015b) show strong inhomogeneities--both north-south, and east-west--in the distribution of condensates in exoplanetary atmospheres. This suggests that the ``patchy" cloud model is highly plausible. 

{  We were able to show analytically why patchy terminator clouds, assuming that the cloudy terminator is dominated by a gray cloud, are able to exactly mimic small scale height atmospheres (typically due to high mean molecular weights) and why the spectral shape of patchy clouds is different than globally uniform clouds.}

 We find that for two well observed planets, HD189733b and HAT-P-11b, that solar composition atmospheres with patchy terminator clouds can match the WFC3 transmission spectra just as well as other interpretations. Unfortunately, the  {  signal-to-noise of} the current WFC3 data for these two objects {  is not high enough to permit a definitive differentiation} between cloud free {  high mean molecular weight} atmospheres, solar composition patchy cloud atmospheres {  with a high altitude ($P_{c}<$ 1 mbar) cloud}, or solar composition {  mid-level (0.1 mbar $<$ P$_{c}$ $<$ 100 mbar)} globally cloudy atmospheres.  This is in contrast to our synthetic example where we could differentiate the non-uniform cloud cover and global cloud coverage (but not high metallicity), unsurprising given the synthetic feature-to-noise ratios are nearly twice as large. Obtaining a similar {\it feature} signal-to-noise for HD189733b and HAT-P-11b should allow us to rule out or confirm the global cloud scenario as parameterized here.  To distinguish the high-metallicity/{  mean molecular weight} case from the solar patchy cloud case one would have to identify high metallicity features at longer wavelengths due to CO and CO$_2$ or {  look at shorter wavelengths as shown in Figure \ref{fig:Figure3}}.  We also found that patchy terminator clouds can result in light curve residuals of up to a couple of hundred ppm, if the transit timing is known to seconds. 
 
 {  In addition to these two differentiating observations one may also use prior information such as interior structure model estimates of the bulk metallicity given the measured mass and radius. However, for highly irradiated hot-Jupiters, such as HD189733b, this task is difficult because there exists a degeneracy between the bulk metal content and the yet-to-be definitively determined  ``inflation" mechanisms (e.g., Batygin, Stevenson, \& Bodenheimer 2011; Miller \& Fortney 2011; Thorngren, Fortney, \& Lopez 2015), and it is not yet clear how the bulk metal content partitions itself between the core and envelope.}  

One may question the plausibility of non-uniform cloud cover on cooler planets. Cooler planets, in general, are expected to have more homogenous day-night temperature contrasts effectively removing night-side cold traps in which classical equilibrium condensates can form. However, we can envision a variety of scenarios in which this may not necessarily matter. For instance, photochemical pre-cursers could be produced on the dayside of the planet, but could continue to polymerize through radical transport on the nightside without being further destroyed by the UV-flux.  Furthermore, Kataria et al. (2014) showed that day-night contrasts can reach up to a few hundred Kelvin at low pressures on warm Neptune-like planets (e.g., GJ1214b), typically where photochemical hazes are formed (e.g., Morley et al. 2013;15).  Additionally, most inhomogeneous cloud cover in the atmospheres of solar system bodies is driven by meridional temperature gradients due to the north-south insolation gradient rather than day-to-night temperature gradients. In essence, {  theoretically}, patchy cloud terminators in transiting planets {  can} be common. 

Another interesting aspect of the 3D problem is that of scale height variations from one terminator to the next in highly irradiated planets, as explored in Fortney et al. (2010) and Burrows et al. (2010). Could variations in scale height be degenerate with non-uniform cloud cover?  If we assume a morning-evening temperature difference of 500 K, a reasonable temperature gradient, we find only 3-5\% fractional annulus area change compared to a half-cloudy half-clear model which gives a 5-11\% fractional area difference.  We should be mindful of this possible additional degeneracy as the larger scale height will occur on the evening terminator of the planet whereas cloud formation will likely occur on the morning. The scale height variation may potentially damp the amplitude of variation due to the patchy clouds by up to half.  

Finally, in order to fully address the complicated nature of clouds, future investigations should consider the full 3-dimensional aspect of the problem and a broader range of cloud prescriptions (as opposed to the simple cloud-top-pressure or uniform gray absorber parameterizations).  Computing 3D transmission spectra from GCM's with clouds of varying particle sizes/compositions will offer more insight into the possible complications when interpreting transit transmission spectra.  Future higher signal to noise observations covering a broader range of wavelengths will likely be able to spectroscopically decipher the differences between high metallicity, uniform cloud cover, and patchy clouds.  Finally, looking at an ensemble of planets over a wide range of temperatures will also aid in deciphering the role of non-uniform cloud cover, similar to what has been done with brown dwarfs. 

%different cloud compositions at terminators

%-show extended wavelengths
%-still in a data regime when results dependent upon assumptions--e.g., high C/O madhu, complicated 1D clouds benneke, patchy clouds or high metallicity-this work

%High C/O is a subset of the cloudy models.  Complicated 1D models and simple 2D models are orthogonal. Real answer is probably some linear combination of those....or really full 3D.
%Looking at these objects in a 1D perspective may not be appropriate

\section{Acknowledgements}	
We thank Jonathan Fortney and Mark Swain for reading the manuscript and providing useful comments;  Daniel Thorngren, Leslie Rogers, \& Caroline Morley for useful discussions.  M.R.L. acknowledges support provided by NASA through Hubble Fellowship grant 51362 awarded by the Space Telescope Science Institute, which is operated by the Association of Universities for Research in Astronomy, In., for NASA, under the contract NAS 5-26555.  V.P. acknowledges support from the Sagan Postdoctoral Fellowship through the NASA Exoplanet Science Institute. The simulations for this research were carried out on the UCSC Hyades computing cluster, which is supported by National Science Foundation (award number AST-1229745) and University of California, Santa Cruz.

\end{document}